\documentclass[conference]{IEEEtran}
\IEEEoverridecommandlockouts

\usepackage{cite}
\usepackage{amsmath,amssymb,amsfonts}
\usepackage{algorithmic}
\usepackage{graphicx}
\usepackage{textcomp}
\usepackage{xcolor}
\usepackage{hyperref}

\usepackage[capitalize]{cleveref}
\usepackage{multirow}
\usepackage{makecell}
\usepackage{mathtools}
\usepackage{enumitem}
\usepackage[strings]{underscore}
\usepackage{url}

\def\BibTeX{{\rm B\kern-.05em{\sc i\kern-.025em b}\kern-.08em
    T\kern-.1667em\lower.7ex\hbox{E}\kern-.125emX}}

\newcommand{\model}[1]{DisCoder}

\begin{document}

\title{High-Fidelity Music Vocoder using\\Neural Audio Codecs}

\author{\IEEEauthorblockN{Luca A. Lanzendörfer}
\IEEEauthorblockA{
\textit{ETH Zurich}\\
lanzendoerfer@ethz.ch}
\and
\IEEEauthorblockN{Florian Grötschla}
\IEEEauthorblockA{
\textit{ETH Zurich}\\
fgroetschla@ethz.ch}
\and
\IEEEauthorblockN{Michael Ungersböck}
\IEEEauthorblockA{
\textit{ETH Zurich}\\
mungersboeck@ethz.ch}
\and
\IEEEauthorblockN{Roger Wattenhofer}
\IEEEauthorblockA{
\textit{ETH Zurich}\\
wattenhofer@ethz.ch}
}

\maketitle

\begin{abstract} \label{sec:abstract}
While neural vocoders have made significant progress in high-fidelity speech synthesis, their application on polyphonic music has remained underexplored. In this work, we propose \model{}, a neural vocoder that leverages a generative adversarial encoder-decoder architecture informed by a neural audio codec to reconstruct high-fidelity 44.1 kHz audio from mel spectrograms. Our approach first transforms the mel spectrogram into a lower-dimensional representation aligned with the Descript Audio Codec (DAC) latent space before reconstructing it to an audio signal using a fine-tuned DAC decoder. \model{} achieves state-of-the-art performance in music synthesis on several objective metrics and in a MUSHRA listening study. Our approach also shows competitive performance in speech synthesis, highlighting its potential as a universal vocoder.
\end{abstract}


\section{Introduction} \label{sec:intro}
Recent developments in audio synthesis have significantly improved the fidelity and efficiency of audio generation systems, especially for text-to-speech applications. While some approaches synthesize waveforms from text directly~\cite{van2016wavenet,oord2018parallel}, most rely on predicting intermediate lower-resolution representations such as mel spectrograms to better manage the complexity of raw audio. Given a mel spectrogram, a vocoder is then used to reconstruct the audio signal represented by the given mel spectrogram. Two notable examples are MelGAN~\cite{kumar2019melgan} and HiFi-GAN~\cite{HiFiGAN}, which use Generative Adversarial Networks (GANs) to efficiently convert mel spectrograms into high-fidelity audio. Using methods such as periodic pattern modeling and anti-aliasing, these models demonstrated notable improvements in computational efficiency and audio synthesis quality. 

Although these developments highlight the potential of vocoders for speech synthesis, which typically features a dominant singular voice, they remain underexplored for polyphonic music generation. Vocoders synthesizing mel spectrograms containing music must handle overlapping instruments, complex vocals, and a significantly broader frequency spectrum. Addressing these challenges would enable high-fidelity audio for a broader set of audio generation models, including text-to-music approaches. There have been various attempts at music synthesis using existing vocoders; however, the quality of the synthesized audio has remained sub-par compared to other approaches~\cite{Forsgren_Martiros_2022,AudioLDM,di2022mel}. To this end, we explore vocoders primarily in the context of polyphonic music synthesis.

In this work, we introduce \model{}, a novel generative adversarial encoder-decoder architecture. Our approach combines ideas from existing vocoders as well as recent developments in neural audio codecs to obtain state-of-the-art performance in music synthesis. By first projecting the input mel spectrogram onto a lower-dimensional space, the \model{} encoder learns to extract the most salient aspects of the audio. The \model{} decoder is initialized from a pre-trained Descript Audio Codec (DAC)~\cite{DAC} decoder and then further fine-tuned, allowing us to leverage the existing prior for high-fidelity audio reconstruction.

\noindent Our contributions can be summarized as follows:

\begin{itemize}[leftmargin=*]
    \item We propose \model{}, a 430M parameter encoder-decoder neural vocoder capable of high-fidelity 44.1 kHz polyphonic music synthesis. Furthermore, we make the codebase and model checkpoints publicly available.
    \item We conduct an ablation study to identify the optimal encoding target in the latent space compared to the number of encoding parameters. Our findings highlight that using the quantized continuous latent space representation of DAC as the prediction target better captures nuances of complex audio signals.
    \item We compare \model{} with previous state-of-the-art and show superior performance in music synthesis across several perceptual metrics and in a MUSHRA listening test.
\end{itemize}

\noindent \textbf{Samples are available online.}\footnote{\url{https://lucala.github.io/discoder/}}

\begin{figure*}[ht!]
  \centering
  \centerline{\includegraphics[width=0.91\textwidth]{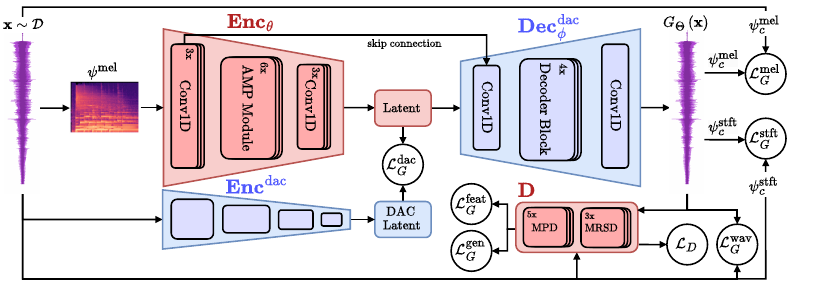}}
    \caption{Proposed \model{} architecture. The mel spectrogram is encoded into a low-dimensional latent space before being decoded to a 44.1 kHz waveform. During the first stage of training, the latent space is aligned with the DAC prior. During the second stage of training this constraint is removed, and a skip connection is introduced to preserve information encoded in the inital mel spectrogram.}
\label{fig:architecture}
\vspace{-8pt}
\end{figure*}

\section{Related Work} \label{sec:related-work}

\textbf{Neural Vocoders.} There has been a long line of work on synthesizing realistic audio signals. The predominant approach has been to leverage a two-stage synthesis by first generating an intermediate representation such as a mel spectrogram, and then synthesizing the audio waveform given the intermediate representation. The Griffin-Lim algorithm~\cite{griffin1984signal} was an early approach to synthesize audio by estimating the phase from a magnitude spectrogram. However, due to compounding errors in the mel spectrogram inversion, this often lead the generated audio to sound unnatural~\cite{wang2017tacotron}. More recently, there have been various approaches inspired by GANs. MelGAN~\cite{kumar2019melgan} generates audio waveforms using a stack of transposed convolution layers, together with multi-scale discriminators for better quality and efficiency in real-time applications. HiFi-GAN~\cite{HiFiGAN} outperformed MelGAN by using a generator with transposed convolutions and residual blocks, combined with multi-scale and multi-period discriminators. BigVGAN~\cite{BigVGAN} and the concurrent work BigVGAN-v2 are able to synthesize high-fidelity 44.1 kHz audio by leveraging a scaled-up architecture with periodic activations and anti-aliased representations.

\textbf{Neural Audio Codecs.} While advancements in neural vocoding have significantly enhanced audio synthesis, similar progress has been observed in the domain of neural audio codecs. SoundStream~\cite{zeghidour2021soundstream} introduced the residual vector quantization to encode audio into a highly compressed discrete latent space. EnCodec~\cite{EnCodec} builds on SoundStream by using a similar encoder-decoder architecture to learn a low-dimensional quantized latent space using a reconstruction loss, a perceptual loss, and an adversarial loss. Descript Audio Codec (DAC)~\cite{DAC} introduced several enhancements, including a reduction of codebook collapse and a multi-scale mel loss, enabling it to obtain better reconstruction of high-frequencies and reduced perceptual audio artifacts.

\section{Method} \label{sec:method}

\subsection{Architecture}

\model{} follows a GAN-based architecture, where the generator consists of an encoder and decoder to synthesize raw audio from mel spectrograms (cf.~\cref{fig:architecture}). The encoder projects the input to a lower dimensional latent representation using multiple 1D convolutional layers and the AMP module as introduced by BigVGAN \cite{BigVGAN}, which uses low-pass filters for anti-aliasing and dilated convolutions with the snake activation function \cite{snake} to provide a periodic inductive bias. During the initial training phase, we align our latent space with the latent space of a pre-trained DAC encoder. Specifically, we use a version of DAC trained on 44.1 kHz audio using nine codebooks, to ensure that the encoder learns a representation that can be faithfully restored to high-fidelity audio. A DAC decoder subsequently transforms this representation into a raw audio signal. After the initial training phase, we remove the latent space loss and activate a skip connection by taking the output of the first convolution layer in the encoder and applying strided average pooling to adjust the time dimension, aligning it with the input of the decoder. The transformed output is then added to the input of the decoders' first convolution layer. This skip connection slightly improved the results in our experiments, likely because it preserved information encoded in the initial mel spectrogram that was beneficial for the DAC decoder during reconstruction.

We use several discriminators to distinguish real from synthesized audio signals. The multi-period discriminator (MPD), introduced by HiFi-GAN \cite{HiFiGAN}, consists of several sub-discriminators that shape the input signal into a 2D representation based on a specified period. By subsequently applying 2D convolutions, each sub-discriminator attends to different periodic components of the input signal. Additionally, we use a multi-resolution spectrogram discriminator (MRSD) introduced by UnivNet~\cite{UnivNet}, where each sub-discriminator operates over spectrograms generated with a different STFT resolution. We use an adaptation from DAC \cite{DAC} that splits the STFT into sub-bands to improve high-frequency components and utilizes both real and imaginary parts to improve phase modeling.

\subsection{Training Losses}
\newcommand{\bx}{\mathbf{x}}
\newcommand{\bq}{\mathbf{q}}
\newcommand{\R}{\mathbb{R}}
\newcommand{\Enc}[1]{\text{Enc}_\theta \left(#1\right)}
\newcommand{\EncDAC}[1]{\text{Enc}^{\text{dac}} \left(#1\right)}
\newcommand{\Dec}[1]{\text{Dec}_\phi^\text{dac} \left(#1\right)}
\newcommand{\Gen}[1]{G_\Theta \left(#1\right)}
\newcommand{\mel}[1]{\psi_c^{\text{mel}} \left( #1 \right)}
\newcommand{\stft}[1]{\psi_c^{\text{stft}} \left( #1 \right)}
\newcommand{\Ex}[1]{\mathbb{E}_{\bx \sim \mathcal{D}}\Bigl[#1\Bigr]}

Our approach uses a variety of loss functions, including reconstruction, adversarial, and feature matching losses, that have been proven effective in previous neural vocoding literature \cite{kumar2019melgan, HiFiGAN, BigVGAN}. An additional latent space alignment loss helps to leverage the DAC prior during the initial phase of training. The generator $\Gen{\bx} \coloneqq \text{Dec}_\phi^\text{dac} \circ \text{Enc}_\theta \circ \psi^\text{mel} \left( \bx \right)$ has parameters $\Theta = \{ \theta, \phi \}$ and takes a normalized audio segment $\bx \in \mathbb{R^{\text{seg}}}$ from our dataset $\mathcal{D}$ as input. The function $\psi^\text{mel}$ converts this signal to a mel spectrogram using a default configuration of 128 mel bins, an FFT and window size of 1024, and a hop length of 256. The waveform loss $\mathcal{L}_G^\text{wav}$ quantifies the difference between the original and reconstructed signal.
\begin{align}
    \mathcal{L}_G^\text{wav}  &= \Ex{\lVert \bx - \Gen{\bx} \rVert_1}
\end{align}
To align the latent space of \model{} with the latent space of the frozen DAC encoder, we use the L1-loss.
\begin{align}
    \mathcal{L}_G^\text{dac}  &= \Ex{\lVert \Enc{ \psi^\text{mel}\left(\bx\right)} - \EncDAC{\bx} \rVert_1}
\end{align}
The multi-scale mel loss \cite{DAC} $\mathcal{L}_G^\text{mel}$ calculates the difference between the mel spectrogram of the original and reconstructed audio for multiple mel configurations. $\mathcal{L}_G^\text{stft}$ is defined similarly for the real STFT output. For $\text{f} \in \{\text{mel}, \text{stft}\}$ the losses are defined as
\begin{align}
    \mathcal{L}_G^\text{f}  &= \sum_c \Ex{\lVert \psi_c^{\text{f}} \left( \bx \right) -  \psi_c^{\text{f}} \left( \Gen{\bx} \right) \rVert_1}
\end{align}
Similar to HiFi-GAN \cite{HiFiGAN}, we use the least-squares GAN loss \cite{LS-GAN} for our adversarial objective. We denote the $i$-th sub-discriminator of type $t \in \mathcal{T}$ as $D_i^t$ where $\mathcal{T} = \left\{ \text{MPD}, \text{MRSD} \right\}$. The number of sub-discriminators of type $t$ are represented by $n^t$. As shown in \cref{fig:architecture}, we use $n^\text{MPD}=5$ and $n^\text{MRSD}=3$.
\begin{align}
    \mathcal{L}_G^\text{gen}  &= \sum_{t \in \mathcal{T}} \sum_{i=1}^{n^t} \Ex{\left(1 - D_i^\text{t}\left(\Gen{\bx}\right)\right)^2} \\
    \mathcal{L}_D               &= \sum_{t \in \mathcal{T}} \sum_{i=1}^{n^t} \Ex{D_i^t\left(\Gen{\bx}\right)^2 + \left(1 - D_i^t\left(\bx\right) \right)^2}
\end{align}
We also use a feature matching loss \cite{feature-matching} $\mathcal{L}_G^\text{feat}$ that minimizes the L1 distance between the feature maps of all sub-discriminators for the original and reconstructed audio. We denote the $j$-th layer of the $i$-th sub-discriminator of type $t$ as $D_i^{t,j}$. Let $L^t$ denote the number of layers and $N_j^t$ the number of features in the $j-$th layer of a discriminator of type $t$. 
\begin{small}
\begin{align}
    \mathcal{L}_G^\text{feat} &= \sum_{t \in \mathcal{T}} \sum_{i=1}^{n^t} \sum_{j=1}^{L^t} \frac{1}{N^t_j} \Ex{\lVert D_i^{t,j}\left(\bx\right) - D_i^{t,j}\left(\Gen{\bx}\right) \rVert_1}
\end{align}
\end{small}
Finally, the generator losses $\mathcal{L}_G^s$ are weighted with $\lambda_s$ for $s \in \left\{ \text{wav}, \text{dac}, \text{mel}, \text{stft}, \text{gen}, \text{feat} \right\}$ and aggregated.

\section{Experiments} \label{sec:experiments}
\subsection{Training}
For training, we convert the MTG-Jamendo~\cite{Jamendo} music and LibriTTS~\cite{LibriTTS} speech datasets to 44.1 kHz mono audio and perform normalization on a per-file basis. Segments of 16,384 frames (0.37s) are randomly extracted and converted to mel spectrograms with 128 mel bins, using an FFT and window size of 1024, and a hop length of 256. We use a learning rate of $\alpha=10^{-4}$ and the AdamW~\cite{AdamW} optimizer with $\beta_1 = 0.8$ and $\beta_2 = 0.99$. In our experiments, exponentially decaying the learning rate with $\gamma = 0.9995$ and clipping gradients above $10^3$ made training significantly more stable. The loss functions are weighted as follows: $\lambda_\text{dac}=\lambda_\text{mel}=15, \lambda_\text{stft}=\lambda_\text{wav}=\lambda_\text{gen}=1, \lambda_\text{feat}=2$. The parameters $\phi$ of the DAC decoder are initially frozen during training to enable the \model{} encoder to approximate the DAC latent space. However, we found that this latent representation could not be reconstructed perfectly, likely due to the missing phase information in the mel spectrogram. By unfreezing the DAC decoder after $10^5$ steps and setting $\lambda_\text{dac}=0$, the fidelity of the reconstructed audio was significantly improved.

\subsection{Validation}

\begin{figure*}[ht!]
    \centering
    \centerline{\includegraphics[width=\textwidth]{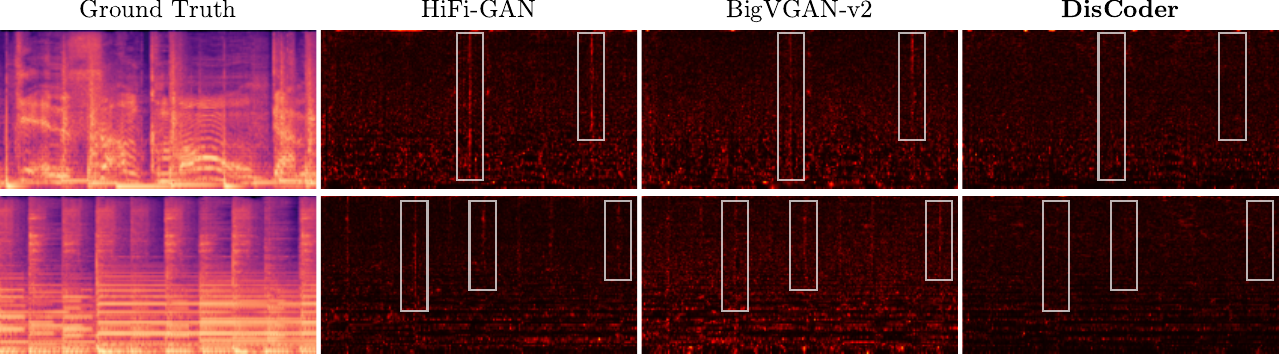}}
    \caption{Comparison of mel spectrogram reconstruction quality between HiFi-GAN, BigVGAN-v2, and \model{} against the ground truth. The three model columns show the absolute error between the mel spectrogram of the synthesized audio and ground truth audio. The rows represent two unseen music clips from the MTG-Jamendo dataset. \model{} contains significantly less pronounced errors compared to HiFi-GAN and BigVGAN-v2.}
    \label{fig:mels}
    \vspace{-10pt}
\end{figure*}

We evaluate our model using several objective metrics. The multi-resolution STFT (MR-STFT)~\cite{MR-STFT} metric aggregates the STFT loss over several configurations that differ in their FFT size, window size, and hop length. The multi-resolution mel spectrogram loss (MR-MEL)~\cite{DAC} computes the sum over the differences between the mel spectrogram of the original and reconstructed audio signal for multiple STFT and mel bin configurations. We also use ViSQOL~\cite{ViSQOL}, a metric that mimics the subjective assessments of audio quality, and CDPAM \cite{CDPAM}, which computes the similarity between two audio signals. For validation on speech, we additionally employ PESQ~\cite{PESQ}. Finally, we conduct a listening test using webMUSHRA~\cite{schoeffler2018webmushra}, following the standard MUSHRA protocol~\cite{MUSHRA}. We present audio experts with 5-second snippets from MTG-Jamendo and MUSDB-HQ~\cite{MUSDB18-HQ} that were not seen during training. The listeners rate these snippets on a scale between 0 and 100 in a double-blind setting. A hidden reference and a 7 kHz low-pass version of the reference are used as anchors.

\begin{table}
\caption{Ablation study comparing prediction targets and model sizes. Results of unseen audio clips from the MTG-Jamendo dataset. The reported values (mean $\pm$ standard deviation) show that predicting the quantized continuous latent Z performs best. }
\begin{center}
\addtolength{\tabcolsep}{-0.4em}
\begin{tabular}{ |cc|cccc| }
    \hline
    Target & Param. &   MR-STFT $\downarrow$ &     MR-MEL $\downarrow$ &      CDPAM $\downarrow$ &       ViSQOL $\uparrow$ \\ 
    \hline
    QL     & 220M &         1.062$\pm$0.08  &         2.768$\pm$0.28  &         0.315$\pm$0.23  &         4.394$\pm$0.20  \\ 
    QL     & 430M &         0.994$\pm$0.08  &         2.577$\pm$0.30  &         0.313$\pm$0.24  &         4.479$\pm$0.19  \\ 
     Z     & 220M &         1.053$\pm$0.09  &         2.625$\pm$0.29  &         0.319$\pm$0.22  &         4.401$\pm$0.21  \\
     Z     & 430M & \textbf{0.943$\pm$0.10} & \textbf{2.456$\pm$0.31} & \textbf{0.312$\pm$0.23} & \textbf{4.512$\pm$0.17} \\ 
    \hline
\end{tabular}
\label{tab:ablation}
\end{center}
\vspace{-10pt}
\end{table}

\subsection{Ablation}
We conducted an ablation study to determine how the number of parameters and different DAC prediction targets impact the audio reconstruction quality. The prediction target determines which latent representation of DAC is learned during the initial training phase and how $\mathcal{L}^\text{dac}_G$ is calculated. As shown in \cref{fig:architecture}, this choice also specifies the bottleneck dimension of our architecture. We used a DAC model trained on 44.1 kHz audio with 9 codebooks, each with 1024 quantized vectors of dimension 8. We tested the following targets:
\begin{itemize}[leftmargin=*]
\item \textbf{Quantized Latents (QL):} In the QL setup, we predict for each audio frame (86 per second) the quantized vectors for all 9 codebooks. The bottleneck dimension of our encoder is therefore $[8\cdot9, \text{frames}]$.
\item \textbf{Quantized continuous representation (Z):} DAC projects the quantized latents to the quantized continuous representation of dimension $[1024, \text{frames}]$ before passing it to the DAC decoder. Using this prediction target, the bottleneck dimension is significantly higher.
\end{itemize}

\begin{table}[t!]

\caption{Results (mean $\pm$ std.) of unseen LibriTTS clips from the \texttt{test-clean} and \texttt{test-other} subsets. \model{} achieves competitive performance on speech reconstruction.}
\vspace{-15pt}
\begin{center}
\addtolength{\tabcolsep}{-0.4em}
\resizebox{\columnwidth}{!}{
    \begin{tabular}{ |l|ccccc| }
        \hline
        Model        &    MR-STFT $\downarrow$ &     MR-MEL $\downarrow$ &      CDPAM $\downarrow$ &       ViSQOL $\uparrow$ &         PESQ $\uparrow$ \\
        \hline
        HiFi-GAN     &         0.886$\pm$0.07  &         2.295$\pm$0.20  &         0.099$\pm$0.05  &         4.512$\pm$0.08  &         3.651$\pm$0.30  \\
        BigVGAN      &         0.802$\pm$0.08  & \textbf{1.819$\pm$0.13} &         0.051$\pm$0.03  &         4.613$\pm$0.07  & \textbf{4.251$\pm$0.15} \\
        BigVGAN-v2   &         0.713$\pm$0.07  &         1.845$\pm$0.14  &         0.053$\pm$0.04  & \textbf{4.691$\pm$0.02} &         4.130$\pm$0.21  \\
        \model{}     & \textbf{0.712$\pm$0.09} &         1.826$\pm$0.15  & \textbf{0.047$\pm$0.03} &         4.664$\pm$0.03  &         4.025$\pm$0.30  \\
        \hline
    \end{tabular}
}
\end{center}
\label{tab:results-speech}
\vspace{-14pt}
\end{table}

We also investigated a third approach, in which the model only predicted the indices of the quantized vectors used for each codebook, resulting in a bottleneck dimension of $[9, \text{frames}]$. Preliminary results showed that our encoder was unable to approximate this low-dimensional representation, which may be attributed to the absent phase information in mel spectrograms. \cref{tab:ablation} shows the results of this ablation study for two model sizes and prediction targets of QL and Z. Each model was trained for 500k steps and evaluated on unseen audio clips from the MTG-Jamendo dataset. The 430M parameter model and the Z prediction target achieved the best results. This suggests that the higher-dimensional latent representation was beneficial, enabling greater modeling flexibility and reducing the impact of early-stage mispredictions in the latent space.

\newcommand{\weightings}{[$\lambda_\text{wav}, \lambda_\text{dac}, \lambda_\text{mel}, \lambda_\text{stft}, \lambda_\text{gen}, \lambda_\text{feat}$]}

\newcommand{\bftab}{\fontseries{b}\selectfont}

\subsection{Results}
We evaluate our proposed architecture against state-of-the-art neural vocoders on both speech and music datasets, utilizing publicly available checkpoints of the models for comparison. For BigVGAN, we test two different variants: the original model trained on the LibriTTS \cite{LibriTTS} dataset using 100 mel bins and a 22 kHz sampling rate, and a more recent version trained on a broader set of data sources using 128 mel bins and a 44.1 kHz sampling rate. The more recent version showed significant improvements and reduced perceptual artifacts for non-speech audio. We also include a pre-trained checkpoint of HiFi-GAN provided by AudioLDM~\cite{AudioLDM} that was trained on the AudioSet~\cite{AudioSet} dataset using 256 mel bins and a 48 kHz sampling rate. For the \model{} architecture, we used a 430M parameter model with the Z prediction target as outlined in \cref{tab:ablation} and extended the training time to 2M steps. 

The mel spectrogram reconstruction quality of two samples from the MTG-Jamendo dataset between HiFi-GAN, BigVGAN-v2, and \model{} is shown in \cref{fig:mels}. Our approach is able to reconstruct music with less artifacts; this is most noticeable when the audio contains sudden modulations or accented beats.

\textbf{Speech.} We evaluate the models on speech clips from LibriTTS~\cite{LibriTTS}. \cref{tab:results-speech} shows the objective metrics, with our architecture and both BigVGAN versions performing similarly, while HiFi-GAN shows slightly worse performance.

\textbf{Music.} \cref{tab:results-music} outlines the validation results of audio clips from MTG-Jamendo~\cite{Jamendo} and MUSDB-HQ~\cite{MUSDB18-HQ}. Except for ViSQOL, \model{} outperforms the previous approaches on objective metrics. In the conducted MUSHRA test, nine participants with a background in audio signal processing were presented with ten randomly selected five-second stimuli. The hidden reference anchor scored the best at $99.78\pm 1.23$, the hidden 7 kHz anchor received the lowest score at $43.53 \pm 17.84$. A Wilcoxon signed rank-test found the difference between the models to be statistically significant (p-value $<0.05$) with \model{} performing the best, and participants noting better reconstruction of pronounced beats matching the results observed in \cref{fig:mels}.

\begin{table}[t!]

\caption{Results (mean $\pm$ std.) of unseen MTG-Jamendo and unseen MUSDB-HQ audio clips. \model{} statistically significantly outperforms other approaches on music synthesis in MUSHRA.}
\vspace{-15pt}
\begin{center}
\addtolength{\tabcolsep}{-0.4em}
\resizebox{\columnwidth}{!}{
    \begin{tabular}{ |l|ccccc| }
        \hline
        Model        &    MR-STFT $\downarrow$ &     MR-MEL $\downarrow$ &      CDPAM $\downarrow$ &      ViSQOL $\uparrow$ &       MUSHRA $\uparrow$  \\
        \hline
        HiFi-GAN     &         0.982$\pm$0.05  &         2.485$\pm$0.24  &         0.073$\pm$0.04  &        4.621$\pm$0.08  &         78.97$\pm$17.12  \\
        BigVGAN      &         1.056$\pm$0.07  &         2.568$\pm$0.23  &         0.073$\pm$0.04  &        4.619$\pm$0.07  &         55.71$\pm$22.12  \\
        BigVGAN-v2   &         0.979$\pm$0.06  &         2.642$\pm$0.27  &         0.080$\pm$0.04  &\textbf{4.641$\pm$0.06} &         83.42$\pm$14.93  \\
        \model{}     & \textbf{0.877$\pm$0.09} & \textbf{2.328$\pm$0.22} & \textbf{0.067$\pm$0.04} &        4.594$\pm$0.10  & \textbf{88.14$\pm$13.91} \\
        \hline
    \end{tabular}
}
\label{tab:results-music}
\vspace{-14pt}
\end{center}
\end{table}

\section{Conclusion} \label{sec:conclusion}
We propose a novel neural vocoder using a discriminative encoder-decoder architecture informed by neural audio codecs for high-fidelity audio synthesis from mel spectrograms. \model{} achieves state-of-the-art performance in music synthesis by connecting the previously distinct areas of neural audio codecs and neural vocoders. 
We validate our approach on several objective metrics and through a MUSHRA test with audio experts. 
We believe this work contributes to the field of audio synthesis and opens up new possibilities for audio generation in various applications, including spectrogram-based generation of high-fidelity polyphonic music.

\clearpage

\bibliographystyle{IEEEtran}
\bibliography{refs}

\begin{thebibliography}{10}
\providecommand{\url}[1]{#1}
\csname url@samestyle\endcsname
\providecommand{\newblock}{\relax}
\providecommand{\bibinfo}[2]{#2}
\providecommand{\BIBentrySTDinterwordspacing}{\spaceskip=0pt\relax}
\providecommand{\BIBentryALTinterwordstretchfactor}{4}
\providecommand{\BIBentryALTinterwordspacing}{\spaceskip=\fontdimen2\font plus
\BIBentryALTinterwordstretchfactor\fontdimen3\font minus \fontdimen4\font\relax}
\providecommand{\BIBforeignlanguage}[2]{{%
\expandafter\ifx\csname l@#1\endcsname\relax
\typeout{** WARNING: IEEEtran.bst: No hyphenation pattern has been}%
\typeout{** loaded for the language `#1'. Using the pattern for}%
\typeout{** the default language instead.}%
\else
\language=\csname l@#1\endcsname
\fi
#2}}
\providecommand{\BIBdecl}{\relax}
\BIBdecl

\bibitem{van2016wavenet}
A.~Van Den~Oord, S.~Dieleman, H.~Zen, K.~Simonyan, O.~Vinyals, A.~Graves, N.~Kalchbrenner, A.~Senior, K.~Kavukcuoglu \emph{et~al.}, ``{WaveNet}: A generative model for raw audio,'' \emph{arXiv preprint arXiv:1609.03499}, vol.~12, 2016.

\bibitem{oord2018parallel}
A.~Oord, Y.~Li, I.~Babuschkin, K.~Simonyan, O.~Vinyals, K.~Kavukcuoglu, G.~Driessche, E.~Lockhart, L.~Cobo, F.~Stimberg \emph{et~al.}, ``Parallel {WaveNet}: Fast high-fidelity speech synthesis,'' in \emph{International conference on machine learning}.\hskip 1em plus 0.5em minus 0.4em\relax PMLR, 2018, pp. 3918--3926.

\bibitem{kumar2019melgan}
K.~Kumar, R.~Kumar, T.~De~Boissiere, L.~Gestin, W.~Z. Teoh, J.~Sotelo, A.~De~Brebisson, Y.~Bengio, and A.~C. Courville, ``{MelGAN}: Generative adversarial networks for conditional waveform synthesis,'' \emph{Advances in neural information processing systems}, vol.~32, 2019.

\bibitem{HiFiGAN}
\BIBentryALTinterwordspacing
J.~Kong, J.~Kim, and J.~Bae, ``{HiFi-GAN}: Generative adversarial networks for efficient and high fidelity speech synthesis,'' \emph{Advances in neural information processing systems}, vol.~33, pp. 17\,022--17\,033, 2020. [Online]. Available: \url{https://proceedings.neurips.cc/paper_files/paper/2020/file/c5d736809766d46260d816d8dbc9eb44-Paper.pdf}
\BIBentrySTDinterwordspacing

\bibitem{Forsgren_Martiros_2022}
\BIBentryALTinterwordspacing
S.~Forsgren and H.~Martiros, ``{Riffusion - Stable diffusion for real-time music generation},'' 2022. [Online]. Available: \url{https://riffusion.com/about}
\BIBentrySTDinterwordspacing

\bibitem{AudioLDM}
\BIBentryALTinterwordspacing
H.~Liu, Z.~Chen, Y.~Yuan, X.~Mei, X.~Liu, D.~Mandic, W.~Wang, and M.~D. Plumbley, ``{AudioLDM}: Text-to-audio generation with latent diffusion models,'' \emph{Proceedings of the International Conference on Machine Learning}, pp. 21\,450--21\,474, 2023. [Online]. Available: \url{https://proceedings.mlr.press/v202/liu23f.html}
\BIBentrySTDinterwordspacing

\bibitem{di2022mel}
B.~Di~Giorgi, M.~Levy, and R.~Sharp, ``Mel spectrogram inversion with stable pitch,'' \emph{arXiv preprint arXiv:2208.12782}, 2022.

\bibitem{DAC}
\BIBentryALTinterwordspacing
R.~Kumar, P.~Seetharaman, A.~Luebs, I.~Kumar, and K.~Kumar, ``High-fidelity audio compression with improved {RVQGAN},'' \emph{Advances in Neural Information Processing Systems}, vol.~36, 2024. [Online]. Available: \url{https://proceedings.neurips.cc/paper_files/paper/2023/file/58d0e78cf042af5876e12661087bea12-Paper-Conference.pdf}
\BIBentrySTDinterwordspacing

\bibitem{griffin1984signal}
D.~Griffin and J.~Lim, ``Signal estimation from modified short-time fourier transform,'' \emph{IEEE Transactions on acoustics, speech, and signal processing}, vol.~32, no.~2, pp. 236--243, 1984.

\bibitem{wang2017tacotron}
Y.~Wang, R.~Skerry-Ryan, D.~Stanton, Y.~Wu, R.~J. Weiss, N.~Jaitly, Z.~Yang, Y.~Xiao, Z.~Chen, S.~Bengio \emph{et~al.}, ``Tacotron: Towards end-to-end speech synthesis,'' \emph{arXiv preprint arXiv:1703.10135}, 2017.

\bibitem{BigVGAN}
\BIBentryALTinterwordspacing
S.-g. Lee, W.~Ping, B.~Ginsburg, B.~Catanzaro, and S.~Yoon, ``{BigVGAN}: A universal neural vocoder with large-scale training,'' \emph{arXiv preprint arXiv:2206.04658}, 2022. [Online]. Available: \url{https://doi.org/10.48550/arXiv.2206.04658}
\BIBentrySTDinterwordspacing

\bibitem{zeghidour2021soundstream}
N.~Zeghidour, A.~Luebs, A.~Omran, J.~Skoglund, and M.~Tagliasacchi, ``{SoundStream}: An end-to-end neural audio codec,'' \emph{IEEE/ACM Transactions on Audio, Speech, and Language Processing}, vol.~30, pp. 495--507, 2021.

\bibitem{EnCodec}
\BIBentryALTinterwordspacing
A.~D{\'e}fossez, J.~Copet, G.~Synnaeve, and Y.~Adi, ``High fidelity neural audio compression,'' \emph{arXiv preprint arXiv:2210.13438}, 2022. [Online]. Available: \url{https://doi.org/10.48550/arXiv.2210.13438}
\BIBentrySTDinterwordspacing

\bibitem{snake}
L.~Ziyin, T.~Hartwig, and M.~Ueda, ``Neural networks fail to learn periodic functions and how to fix it,'' \emph{Advances in Neural Information Processing Systems}, vol.~33, pp. 1583--1594, 2020.

\bibitem{UnivNet}
\BIBentryALTinterwordspacing
W.~Jang, D.~Lim, J.~Yoon, B.~Kim, and J.~Kim, ``{UnivNet}: A neural vocoder with multi-resolution spectrogram discriminators for high-fidelity waveform generation,'' in \emph{Proc. Interspeech 2021}, 2021, pp. 2207--2211. [Online]. Available: \url{https://doi.org/10.21437/Interspeech.2021-1016}
\BIBentrySTDinterwordspacing

\bibitem{LS-GAN}
\BIBentryALTinterwordspacing
X.~Mao, Q.~Li, H.~Xie, R.~Y. Lau, Z.~Wang, and S.~Paul~Smolley, ``Least squares generative adversarial networks,'' in \emph{Proceedings of the IEEE international conference on computer vision}, 2017, pp. 2794--2802. [Online]. Available: \url{https://doi.org/10.1109/ICCV.2017.304}
\BIBentrySTDinterwordspacing

\bibitem{feature-matching}
A.~B.~L. Larsen, S.~K. S{\o}nderby, H.~Larochelle, and O.~Winther, ``Autoencoding beyond pixels using a learned similarity metric,'' in \emph{International conference on machine learning}.\hskip 1em plus 0.5em minus 0.4em\relax PMLR, 2016, pp. 1558--1566.

\bibitem{Jamendo}
\BIBentryALTinterwordspacing
D.~Bogdanov, M.~Won, P.~Tovstogan, A.~Porter, and X.~Serra, ``The {MTG-Jamendo} dataset for automatic music tagging,'' in \emph{Machine Learning for Music Discovery Workshop, International Conference on Machine Learning (ICML 2019)}, Long Beach, CA, United States, 2019. [Online]. Available: \url{http://hdl.handle.net/10230/42015}
\BIBentrySTDinterwordspacing

\bibitem{LibriTTS}
\BIBentryALTinterwordspacing
H.~Zen, V.~Dang, R.~Clark, Y.~Zhang, R.~J. Weiss, Y.~Jia, Z.~Chen, and Y.~Wu, ``{LibriTTS}: A corpus derived from {LibriSpeech} for text-to-speech,'' in \emph{Proc. Interspeech 2019}, 2019, pp. 1526--1530. [Online]. Available: \url{https://doi.org/10.21437/Interspeech.2019-2441}
\BIBentrySTDinterwordspacing

\bibitem{AdamW}
\BIBentryALTinterwordspacing
I.~Loshchilov and F.~Hutter, ``Decoupled weight decay regularization,'' \emph{arXiv preprint arXiv:1711.05101}, 2017. [Online]. Available: \url{https://doi.org/10.48550/arXiv.1711.05101}
\BIBentrySTDinterwordspacing

\bibitem{MR-STFT}
\BIBentryALTinterwordspacing
R.~Yamamoto, E.~Song, and J.-M. Kim, ``Parallel {WaveGAN}: A fast waveform generation model based on generative adversarial networks with multi-resolution spectrogram,'' in \emph{ICASSP 2020-2020 IEEE International Conference on Acoustics, Speech and Signal Processing (ICASSP)}.\hskip 1em plus 0.5em minus 0.4em\relax IEEE, 2020, pp. 6199--6203. [Online]. Available: \url{https://doi.org/10.1109/ICASSP40776.2020.9053795}
\BIBentrySTDinterwordspacing

\bibitem{ViSQOL}
\BIBentryALTinterwordspacing
M.~Chinen, F.~S. Lim, J.~Skoglund, N.~Gureev, F.~O'Gorman, and A.~Hines, ``{ViSQOL} v3: An open source production ready objective speech and audio metric,'' in \emph{2020 twelfth international conference on quality of multimedia experience (QoMEX)}.\hskip 1em plus 0.5em minus 0.4em\relax IEEE, 2020, pp. 1--6. [Online]. Available: \url{https://doi.org/10.1109/QoMEX48832.2020.9123150}
\BIBentrySTDinterwordspacing

\bibitem{CDPAM}
P.~Manocha, Z.~Jin, R.~Zhang, and A.~Finkelstein, ``{CDPAM}: Contrastive learning for perceptual audio similarity,'' in \emph{ICASSP 2021-2021 IEEE International Conference on Acoustics, Speech and Signal Processing (ICASSP)}.\hskip 1em plus 0.5em minus 0.4em\relax IEEE, 2021, pp. 196--200.

\bibitem{PESQ}
\BIBentryALTinterwordspacing
A.~Rix, J.~Beerends, M.~Hollier, and A.~Hekstra, ``Perceptual evaluation of speech quality ({PESQ})-a new method for speech quality assessment of telephone networks and codecs,'' in \emph{2001 IEEE International Conference on Acoustics, Speech, and Signal Processing. Proceedings (Cat. No.01CH37221)}, vol.~2, 2001, pp. 749--752 vol.2. [Online]. Available: \url{https://doi.org/10.1109/ICASSP.2001.941023}
\BIBentrySTDinterwordspacing

\bibitem{schoeffler2018webmushra}
M.~Schoeffler, S.~Bartoschek, F.-R. St{\"o}ter, M.~Roess, S.~Westphal, B.~Edler, and J.~Herre, ``{webMUSHRA} — a comprehensive framework for web-based listening tests,'' 2018.

\bibitem{MUSHRA}
\BIBentryALTinterwordspacing
I.~Recommendation \emph{et~al.}, ``Method for the subjective assessment of intermediate sound quality ({MUSHRA}),'' \emph{ITU, BS}, pp. 1543--1, 2001. [Online]. Available: \url{https://www.itu.int/dms_pubrec/itu-r/rec/bs/R-REC-BS.1534-3-201510-I!!PDF-E.pdf}
\BIBentrySTDinterwordspacing

\bibitem{MUSDB18-HQ}
\BIBentryALTinterwordspacing
Z.~Rafii, A.~Liutkus, F.-R. Stöter, S.~I. Mimilakis, and R.~Bittner, ``{MUSDB18-HQ} - an uncompressed version of {MUSDB18},'' Aug. 2019. [Online]. Available: \url{https://doi.org/10.5281/zenodo.3338373}
\BIBentrySTDinterwordspacing

\bibitem{AudioSet}
\BIBentryALTinterwordspacing
J.~F. Gemmeke, D.~P.~W. Ellis, D.~Freedman, A.~Jansen, W.~Lawrence, R.~C. Moore, M.~Plakal, and M.~Ritter, ``{Audio Set}: An ontology and human-labeled dataset for audio events,'' in \emph{2017 IEEE International Conference on Acoustics, Speech and Signal Processing (ICASSP)}, 2017, pp. 776--780. [Online]. Available: \url{https://doi.org/10.1109/ICASSP.2017.7952261}
\BIBentrySTDinterwordspacing

\end{thebibliography}

\end{document}